\documentclass[showpacs,preprintnumbers,twocolumn]{revtex4}
\usepackage{amsmath}
\usepackage{graphicx}
\usepackage{bm}

\input{tcilatex}

\begin{document}

\preprint{}
\title{Novel approach to a perfect lens}
\author{V.V. Klimov}
\email{vklim@sci.lebedev.ru}
\affiliation{P.N. Lebedev Physical Institute, Russian Academy of Sciences, 53 Leninsky
Prospekt, Moscow 119991, Russia}
\pacs{78.20.Ci, 42.30.Wb, 73.20.Mf, 78.66.Bz}

\begin{abstract}
Within the framework of an exact analytical solution of Maxwell equations in
a space domain, it is shown that optical scheme based on a slab with
negative refractive index ($n=-1$) (Veselago lens or Pendry lens) does not
possess focusing properties in the usual sense . In fact, the energy in such
systems does not go from object to its "image", but from object and its
"image" to an intersection point inside a metamaterial layer, or vice versa.
A possibility of applying this phenomenon to a creation of entangled states
of two atoms is discussed.
\end{abstract}

\maketitle

Recently due to the work of Veselago \cite{Veselago}, much attention has
been paid to a so called perfect lens, whose properties are due to
particular features of media with negative refractive index. Keen interest
to this theme was aroused by the work of Pendry \cite{Pendry}, where because
of impossibility to produce media with negative refractive index he had
proposed to use metallic films with $\varepsilon =-1$ for near field
focusing. The attempts to make a perfect lens have already involved
interesting experiments on optical near field focusing (superlenses \cite%
{Zhang1}, hyperlenses \cite{Zhang2,Smolyaninov}, and nanolenses \cite{Kawata}%
). However, the Veselago perfect lenses \cite{Veselago} with $n=-1$ have not
yet been achieved, and as will be shown below, they cannot be realized due
to the reasons not at all connected with technical difficulties of making a
metamaterial with negative refractive index. The fact is that a conventional
picture of operation of a perfect lens is incorrect, because it disagrees
with the standard Maxwell equations.

\begin{figure}[tbp]
\includegraphics[width=7.5cm] {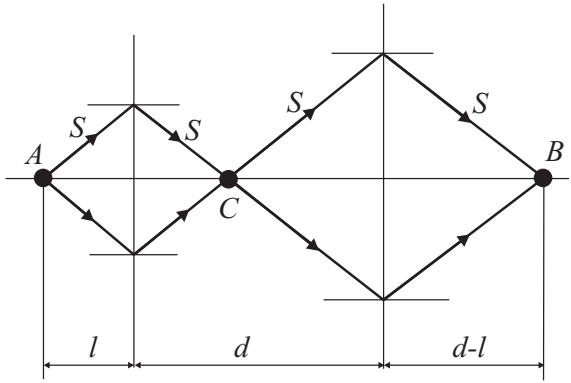}
\caption{Standard ray picture in a perfect lens [1,2]. The arrows show
direction of the energy flow S.}
\label{fig1}
\end{figure}

The principle of operation of a perfect lens is illustrated by a scheme of
Fig.1, which had been first represented in Veselago's work \cite{Veselago}.
In accordance with the Fresnel laws for the interface of media with
refractive index $n=1$ (usual material) and with refractive index $n=-1$
("left-handed" metamaterial), the rays propagating from an object (point A)
intersect at point B thus forming an image. This is an elementary picture,
at first sight, but it becomes non- elementary if we consider spatial
structure of electromagnetic fields at points B and C, and a direction of
energy propagation in such systems. Here a question arises about singular
points of electromagnetic field in homogeneous space, into which the energy
flows in, from one direction, and flows out, from the other.

Consideration based on the field expansion over plane and evanescent waves 
\cite{Pendry} proves to be deeper than a ray approach. It shows that near
field amplification occurs in a metamaterial layer with $n=-1$ due to the
surface plasmon resonance, which is connected with a subsequent perfect
focusing. Such kind of argumentation was put in doubt in \cite{Garcia,Hooft}%
. The case of almost-perfect lens, that is the case of slab with refractive
index $n\approx-1$ was considered in \cite{Merlin}, but no focal spot at
point B was detected. Thus, until the present work, no convincing solution
of that problem had been obtained in a spatial domain, as far as I know.

\begin{figure}[tbp]
\includegraphics[width=7.5cm] {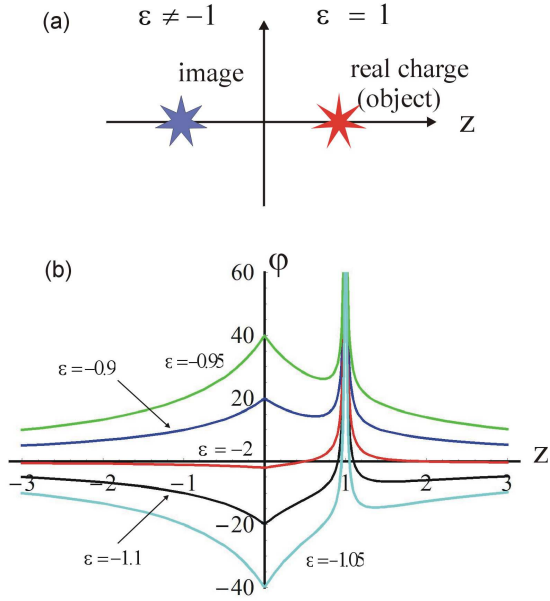}
\caption{Illustration to a solution of quasi-static problem on a charge near
semi-infinite space ; a) geometry of the problem ; b) potential distribution
on the system axis at different values of dielectric permittivity.}
\label{fig2}
\end{figure}

To make that problem clear, we first consider the case of near fields, i.e.,
the quasi-static Pendry perfect lens \cite{Pendry}. Let us analyze a point
charge $q=1$ in vacuum near a half- space with dielectric permittivity $%
\varepsilon $ (Fig.2). Solution of such an electrostatic problem is well
known \cite{Landau}

\begin{eqnarray}
\varphi ^{\left( 1\right) } &=&\frac{1}{\left\vert \mathbf{r}-\mathbf{r}%
_{0}\right\vert }-\frac{\varepsilon -1}{\varepsilon +1}\frac{1}{\left\vert 
\mathbf{r}-\widehat{\mathbf{r}}_{0}\right\vert },z>0  \label{eq1} \\
\varphi ^{\left( 2\right) } &=&\frac{2}{\varepsilon +1}\frac{1}{\left\vert 
\mathbf{r}-\mathbf{r}_{0}\right\vert },z<0  \label{eq2}
\end{eqnarray}

where $\mathbf{r}_{0},\widehat{\mathbf{r}}_{0}=\mathbf{r}_{0}\left(
z\rightarrow -z\right) $ are radius-vectors of real charge and its image
respectively.

From that solution it is immediately seen that as the permittivity tends to
-1, the solution tends to infinity near a surface, and thus it does not
exist within the limit of $\varepsilon =-1$ (Fig.2b). From the physical
viewpoint, the solution tendency to the infinity in whole space is due to
the excitation of resonant surface plasmon waves at the interface.
Meanwhile, the question arises: is there any meaningful solution at $%
\varepsilon =-1$? From the formal mathematical point of view, there is no
bounded solution in whole space (except the charge position point). But if
we admit the presence of only one or several singular points in the region
of metamaterial then the solution becomes possible and obtains the form:

\begin{eqnarray}
\varphi ^{\left( 1\right) } &=&\frac{1}{\left\vert \mathbf{r}-\mathbf{r}%
_{0}\right\vert }+\dsum\limits_{i}Q_{i}\frac{1}{\left\vert \mathbf{r}-%
\widehat{\mathbf{r}}_{i}\right\vert },z>0  \notag \\
\varphi ^{\left( 2\right) } &=&\frac{1}{\left\vert \mathbf{r}-\widehat{%
\mathbf{r}}_{0}\right\vert }+\dsum\limits_{i}Q_{i}\frac{1}{\left\vert 
\mathbf{r}-\mathbf{r}_{i}\right\vert },z<0  \label{eq3}
\end{eqnarray}

where the charges $Q_{i}$ and their \textquotedblleft
positions\textquotedblright\ $\mathbf{r}_{i},\widehat{\mathbf{r}}_{i}=%
\mathbf{r}_{i}\left( z\rightarrow -z\right) $ are arbitrary.

\begin{figure}[tbp]
\includegraphics[width=7.5cm] {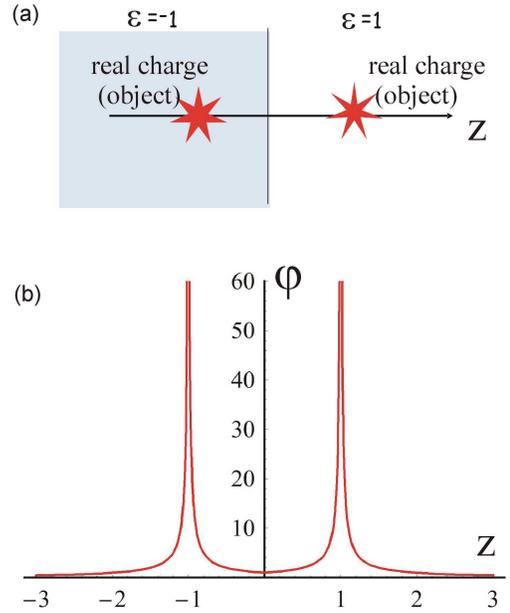}
\caption{Illustrations to the solution of a quasi-static problem
corresponding to the Pendry lens [2]. a) geometry of the problem ; b)
potential distribution on the system axis.}
\label{fig3}
\end{figure}

Thus, if there exist two equal and symmetrically situated charges, then the
Laplace equation with standard boundary conditions at the interface has
quite meaningful (but not the unique ones) solutions. Figure 3a,b
illustrates such solutions for two equal real charges located symmetrically
relative to interface. It is very important that the fields are bounded at
the interface, and the surface plasmons are not excited.

For a finite-thickness layer with $\varepsilon =-1$, the solution has the
analogous form:

\begin{eqnarray}
\varphi ^{\left( 3\right) } &=&\frac{1}{\left\vert \mathbf{r}-\mathbf{r}%
_{C}\right\vert },\text{inside slab}  \notag \\
\varphi ^{\left( 1\right) } &=&\frac{1}{\left\vert \mathbf{r}-\mathbf{r}%
_{A}\right\vert },\text{in the source region}  \label{eq4} \\
\varphi ^{\left( 2\right) } &=&\frac{1}{\left\vert \mathbf{r}-\mathbf{r}%
_{B}\right\vert },\text{in the "image" region}  \notag
\end{eqnarray}%
where the choice of a homogeneous solution is not arbitrary, and the
radius-vectors \ $\mathbf{r}_{A},\mathbf{r}_{B},\mathbf{r}_{C}$ correspond
to the A, B, and C points in Fig.1.

From (\ref{eq3}) and (\ref{eq4}) it is seen that solutions of the Maxwell
equations corresponding to the source A in the region $z>0$ and having
singularities at the ray intersection points B and/or C (Fig.1), and which
correspond to real charges, exist both in the case of a half-space with $%
\varepsilon =-1$ and in the case of slab with $\varepsilon =-1$. However,
unlike a seeming asymmetry between object and its image of Fig.1, all
charges of the system are now symmetric, have the same sign, and each of
them cannot be considered as images of the others.

Thus, by an example of a quasi-static Pendry case \cite{Pendry} it is
already seen that optical system shown in Fig.1 cannot be considered as the
lens, and it must necessarily include three real sources of a charge.

It is difficult to define energy flow in the quasi-static case, and
therefore, the above reasoning is only the indirect evidence that the usual
picture is not correct, and one cannot use the system of Fig.1 in the regime
of a usual lens. For final evidence of that fact, we consider full
electrodynamics problem with one or several interfaces between the
right-handed medium with $\varepsilon =1,\mu =1$ and the left-handed medium
with $\varepsilon =-1,\mu =-1(n=-1)$. One can verify that there is no
solution of the Maxwell equations with one source in the right-hand medium
and without any sources in the "left-handed" half-space. However, one may
derive explicit solution of the Maxwell equations by admitting the existence
of a finite number of singular points, where charges and currents arise.

Consider a case of a single interface between right-and "left-handed"
matter. If the right- handed half-space contains a dipole with
boundary-parallel orientation, then in the left-hand matter the dipole must
be placed, which is directed in the opposite direction at the mirror-
symmetrical point (see Fig.4). Formally speaking, the Hertz vector in the
"right-handed" medium has the form:

\begin{equation}
\mathbf{\Pi }^{\left( 1\right) }=\mathbf{d}_{0}\frac{\exp \left( i\omega
\left\vert \mathbf{r}-\mathbf{r}_{0}\right\vert /c\right) }{\left\vert 
\mathbf{r}-\mathbf{r}_{0}\right\vert }e^{-i\omega t}  \label{eq5}
\end{equation}

and in the "left-handed" medium,

\begin{equation}
\mathbf{\Pi }^{\left( 2\right) }=\mathbf{d}_{0}\frac{\exp \left( i\omega
\left\vert \mathbf{r}-\widehat{\mathbf{r}}_{0}\right\vert /c\right) }{%
\left\vert \mathbf{r}-\widehat{\mathbf{r}}_{0}\right\vert }e^{-i\omega t}
\label{eq6}
\end{equation}

\begin{figure}[tbp]
\includegraphics[width=7.5cm] {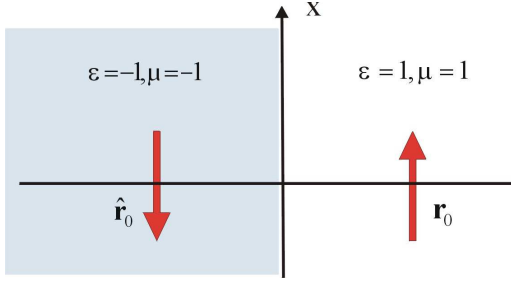}
\caption{Illustration to the solution of the full system of Maxwell
equations for the dipole in the presence of semi-infinite space from the
"left" matter}
\label{fig4}
\end{figure}

Using relations between Hertz potential $\mathbf{\Pi }$ and fields $\mathbf{%
E,H}$\cite{Vainstein}

\begin{eqnarray}
\mathbf{E} &=&\nabla \left( \nabla \Pi \right) +\varepsilon \mu \left(
\omega /c\right) ^{2}\mathbf{\Pi }  \notag \\
\mathbf{H} &=&-i\varepsilon \left( \omega /c\right) \nabla \times \mathbf{%
\Pi }  \label{eq7}
\end{eqnarray}

from (\ref{eq5}) and (\ref{eq6}) we find the electric and magnetic field
fields, which satisfy standard continuity conditions for tangential
components, that should be solution of Maxwell equation. Note that both (\ref%
{eq5}) and (\ref{eq6}) describe phase propagation from the source, and
energy flow from the source ("right-handed" medium) and to the source
("left-handed" medium).

For the "left-handed" slab, the solution is built analogously, and has the
form

\begin{eqnarray}
\mathbf{\Pi }^{\left( 1\right) } &=&\mathbf{d}_{0}\frac{\exp \left( i\omega
\left\vert \mathbf{r}-\mathbf{r}_{A}\right\vert /c\right) }{\left\vert 
\mathbf{r}-\mathbf{r}_{A}\right\vert }e^{-i\omega t},  \notag \\
&&\text{in the source region}  \notag \\
\mathbf{\Pi }^{\left( 3\right) } &=&\mathbf{d}_{0}\frac{\exp \left( i\omega
\left\vert \mathbf{r}-\mathbf{r}_{C}\right\vert /c\right) }{\left\vert 
\mathbf{r}-\mathbf{r}_{C}\right\vert }e^{-i\omega t},  \notag \\
&&\text{inside slab}  \notag \\
\mathbf{\Pi }^{\left( 2\right) } &=&\mathbf{d}_{0}\frac{\exp \left( i\omega
\left\vert \mathbf{r}-\mathbf{r}_{B}\right\vert /c\right) }{\left\vert 
\mathbf{r}-\mathbf{r}_{B}\right\vert }e^{-i\omega t},  \notag \\
&&\text{in the "image" region}  \label{eq8a}
\end{eqnarray}%
where $\mathbf{r}_{A},\mathbf{r}_{B}$, and $\mathbf{r}_{C}$ are
radius-vectors of ray intersection points $A,B,C$ in Fig.1. It is very
interesting that the solution (\ref{eq8a}) remains valid even in the case $%
l>d$ , but in this case positions of $B$ and $C$ are inside and outside
slab, respectively. Of course in that case there are no singularities except
for the dipole source at point $A$ and one cannot speak about lens effect.

The picture for the normally oriented dipole is fully analogous. In that
case, however, one should choose a antisymmetrical combination of Hertz
potentials.

To illustrate physical meaning of our solution and singular points $A,B,C$ ,
we plot Umov-Poynting vector distribution in the y-z plane (normal plane to
the dipole orientation) for the case of parallel-boundary dipole (Fig.5).
For more detailed demonstration see \cite{Klimov4}.

\begin{figure}[tbp]
\includegraphics[width=8cm] {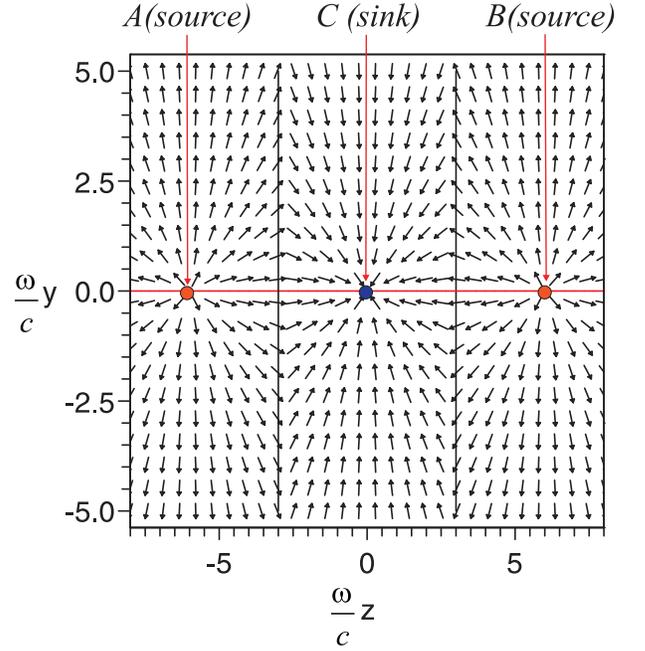}
\caption{Energy flow distribution in the y-z plane that corresponds to the
found solution (8) with dipole along x-axis. Distance from the dipole to the
layer is $l=3$, layer thickness, $d=2l$.}
\label{fig5}
\end{figure}

From Fig.5 (and \cite{Klimov4}) it is seen that energy outgoes from two
sources (A and B) , outside the \textquotedblleft left handed
\textquotedblright slab, and moves towards the point C (inside slab), where
it is absorbed by the third source to be named sink. Obviously, there is
also a solution where the energy propagates from source C in the
"left-handed" slab to the sinks (A and B) in usual matter. Thus, the energy
flow in the system shown in Fig.1 is not correct. Instead of it, one should
use Fig.6, whose picture is radically different from that which prevails
nowadays in the metamaterials community.

\begin{figure}[tbp]
\includegraphics[width=7.5cm] {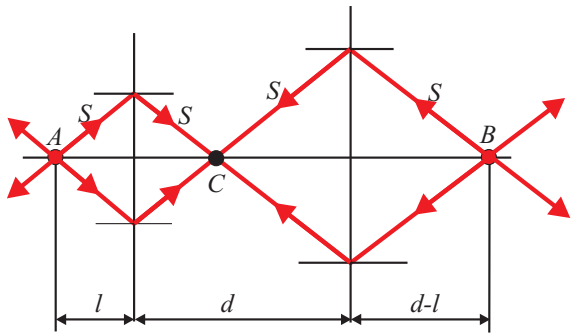}
\caption{Correct ray picture in the presence of the left-handed slab. Arrows
denote direction of the energy flow S.}
\label{fig6}
\end{figure}

Above we have presented the solution in the absence of losses. We have also
found the solution of Maxwell equation with three sources (\ref{eq8a}) for
the case with losses as well, and this solution is continuously transformed
into the solution (\ref{eq8a}) when losses tend to zero. In the case of the
solution we have proposed, the symmetry of the source disposition does not
let the surface plasmon waves to be excited, and the solution, therefore,
remains finite as the losses tend to zero except for $A,B,$ and $C$ points.

On the other hand, there also exists a solution of Maxwell equations with
only one dipole source near the "left-handed" layer with losses, but when $%
l<d$ it tends to infinity as losses are diminishing, because in such a
system there occurs resonant excitation of surface plasmon waves. It is
interesting to note that for $l>d$ this solution tends to the solution (\ref%
{eq8a}) in the case of small losses! In any case, the solution of Maxwell
equations with only one source near the "left-handed" layer with losses has
maximum at the interface only and cannot be considered as conventional lens
with well defined focal spot.

Despite the found solution does not allow us to consider the "left-handed"\
slab as the perfect lens, we believe that it opens up new possibilities of
using the layer with negative refractive index, $n=-1$. For example, by
placing two excited atoms at points A and B and a non- excited one, at point 
$C$, one may excite an atom in point $C$ with a probability close to unit
(the probability that this atom remains unexcited is equal to the
probability that both photons had flown in the direction opposite to the
layer, that is, 1/4). An inverse initial conditions, with one excited atom
(at C point) and two unexcited atoms are also possible. The second situation
may turn out to be even more interesting because in this case there may be
formed entangled state of the excited atoms (at A and B points). Perhaps,
new logical elements for quantum computers can be elaborated on this
direction.

Thus, we have strictly shown that a slab with negative refractive index $%
n=-1 $ could not be considered as a focusing element. We have also proposed
to use more complex configurations of the sources and sinks in order to
reveal new very interesting features of the layer with negative refractive
index.

The author would like to express his gratitude to the Russian Foundation for
Basic Research (grants \# 05-02-19647 and \#07-02-01328) and the Presidium
of the Russian Academy of Sciences for financial support of the present
work. The author is also grateful to N.I. Zheludev for hospitality and
useful discussions of this work.

\end{document}